\documentstyle[emulapj,epsfig]{article}

\begin{document}

\title{The Propeller Regime of Disk
Accretion to a Rapidly Rotating Magnetized Star}

\author{M.M.~Romanova}
\affil{Department of Astronomy, Cornell University, Ithaca, NY
14853-6801;~ romanova@astro.cornell.edu}

\author{G.V.~Ustyugova}
\affil{Keldysh Institute of Applied Mathematics, Russian Academy
of Sciences, Moscow, Russia;~ ustyugg@spp.Keldysh.ru}

\author{A.V.~Koldoba}
\affil{Institute of Mathematical Modelling, Russian Academy of
Sciences, Moscow, Russia;~koldoba@spp.Keldysh.ru}

\author{R.V.E.~Lovelace}
\affil{Department of Astronomy, Cornell University, Ithaca, NY
14853-6801; ~RVL1@cornell.edu }

\begin{abstract}

The propeller regime of disk accretion to a rapidly rotating
magnetized star is investigated here for the first time by
axisymmetric 2.5D magnetohydrodynamic simulations. An expanded,
closed magnetosphere forms in which the magnetic field is
predominantly toroidal. A smaller fraction of the star's poloidal
magnetic flux inflates vertically, forming a magnetically
dominated tower. Matter accumulates in the equatorial region
outside magnetosphere and accretes to the star quasi-periodically
through elongated funnel streams which cause the magnetic field to
reconnect. The star spins-down owing to the interaction of the
closed magnetosphere with the disk. For the considered conditions,
the spin-down torque varies with the angular velocity of the star
($\omega_*$) as $\sim - \omega*^{1.3}$ for fixed mass accretion
rate. The propeller stage may be important in the evolution of
X-ray pulsars, cataclysmic variables and young stars. In
particular, it may explain the
present slow rotation of the
classical T Tauri stars.

\end{abstract}


\section{Introduction}

For a rapidly rotating magnetized star, the centrifugal force at
the magnetospheric radius may be larger than the gravitational
force. In this case the magnetosphere acts as a ``propeller"
pushing away incoming matter (Davidson \& Ostriker 1973;
Illarionov \& Sunyaev 1975; Lipunov 1992). The propeller regime
has been discussed in connection with different astrophysical
situations (e.g., Stella, White, \& Rosner 1986; Treves, Colpi \&
Lipunov 1993; Cui 1997; Alpar 2001; Eksi \& Alpar 2003; Mori \&
Ruderman 2003).

A detailed understanding
of the propeller regime is important
for the analysis of a number of
astrophysical systems, including
Classical T Tauri stars (CTTS),
cataclysmic variables, X-ray
pulsars, isolated neutron stars,
and possibly anomalous X-ray
pulsars.

For quasi-spherical accretion, the propeller regime was
investigated analytically by Davies, Fabian, and Pringle (1979),
Davies and Pringle (1981), and Ikhsanov (2002), and numerically by
Romanova et al. (2003a) for the case of an aligned magnetic
rotator. The numerical simulations have shown that the propelling
star expels significant magnetic flux in the radial direction,
thus forming the low-density, azimuthally wrapped equatorial
magnetic wind. At the same time, some matter was observed to
accrete along the magnetic poles of the star (as observed in the
non-rotating case, Toropina et al. 2003).
These simulations revealed a complicated
flow geometry not considered in
the theoretical models.

 Disk accretion to a star in the propeller regime was
investigated theoretically by Li and Wickramasinghe (1997),
Lovelace, Romanova, and Bisnovatyi-Kogan (1999). Wang and
Robertson (1985) analyzed numerically the disk-star interaction in
the propeller regime, but only in the equatorial plane. Romanova
et al. (2002; hereafter RUKL) did axisymmetric simulations of disk
accretion to the fast rotating star, however only in the regime of
very soft propeller. In this paper we discuss for the first time
axisymmetric MHD simulations of disk accretion to fast propellers.

\begin{figure*}[t]
\epsscale{1.6} \plotone{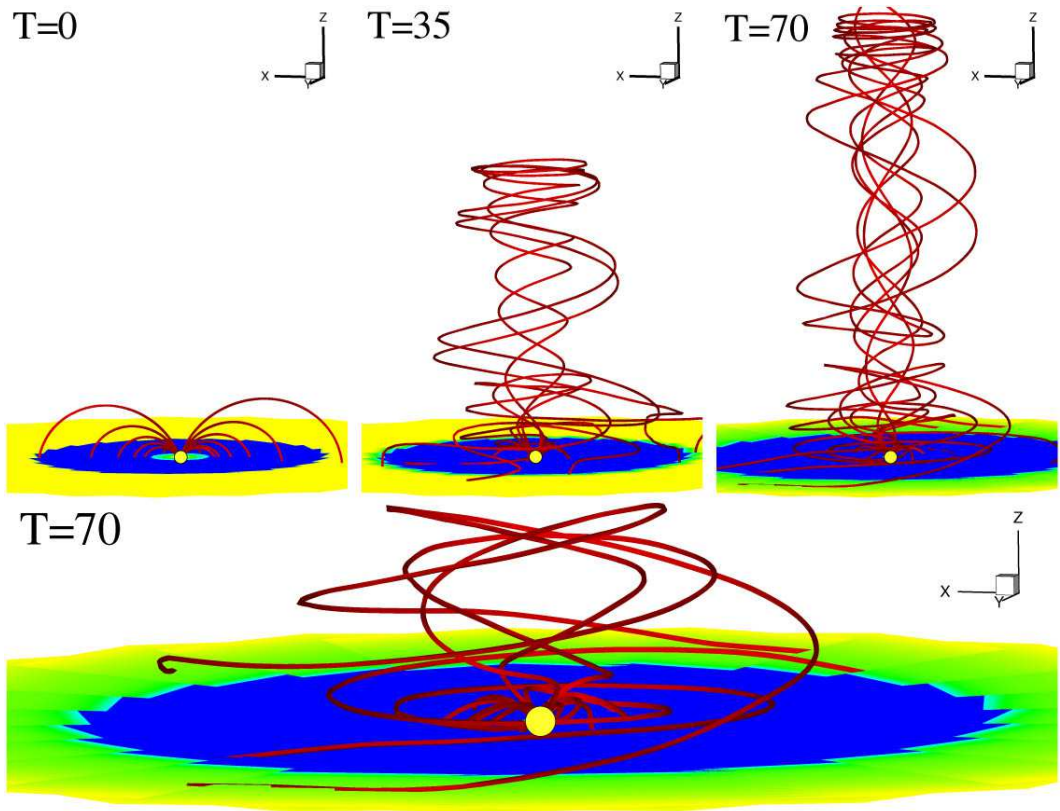} \caption{The figure shows the
opening of the magnetic field lines during the disk-star
interaction at $T=0, 35, 70$. Time $T$ is measured in Keplerian
periods of rotation at $r=1$. The bottom plot represents expanded
view of the magnetosphere near the disk at $T=70$, where field
lines are closed or only partially opened.}
\label{Figure 1}
\end{figure*}

\begin{figure*}[b]
\epsscale{1.6} \plotone{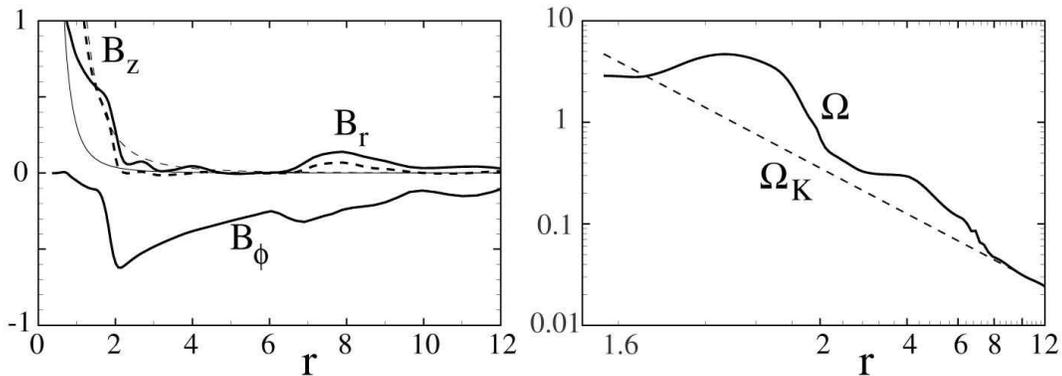} \caption{ Radial variations of the
magnetic field components and the angular velocity above the
equatorial plane of the magnetosphere of the star for
$\omega_*=0.58$ at $T=70$. In the left-hand panel the thick solid
lines show $B_r$ and $B_\phi$ components. The thick dashed line
shows the $B_z$ component. The thin solid and dashed lines show
the initial dipole field $B_z$ and $B_r$. In the right-hand panel
the solid and dashed lines show the logarithm of the angular
velocity and Keplerian angular velocity. } \label{Figure 2}
\end{figure*}

\begin{figure*}[t]
\epsscale{2.} \plotone{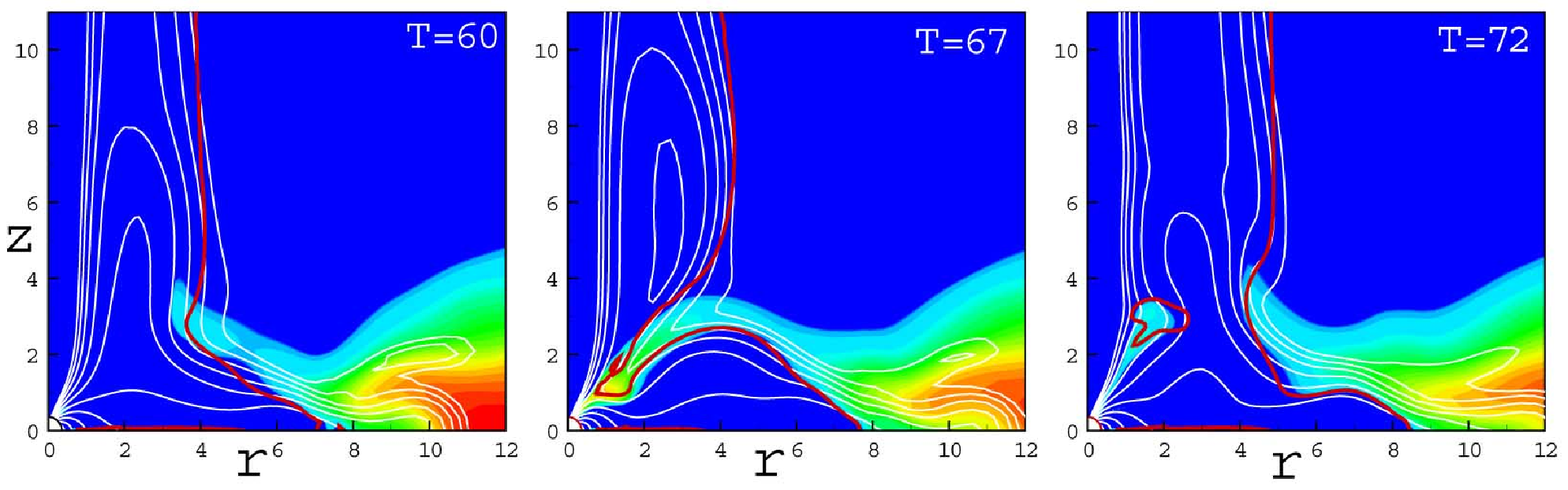} \caption{ Episode of accretion for
$\omega_*=0.58$. From left to right the figure shows before,
during, and after the accretion event. The color background shows
the density levels, that change from $\rho=0.003$ (dark blue) to
$\rho=1.6$ (red). The white lines are magnetic field lines. The
red solid line shows the surface $\beta=E_{matter}/E_{mag} =1$
which separates the magnetically dominated region from the matter
dominated region. The Alfv\'en radius ($\beta=1$) is $r_A\approx
8$. For a non-rotating star, $r_A\approx 1$.} \label{Figure 3}
\end{figure*}

\section{The Model}

Axisymmetric MHD
simulations have been done for
the interaction of an accretion
disk with a rapidly
rotating magnetized star in the propeller regime.
The arrangement of the model is similar to that used in RUKL for
analysis of disk accretion to slowly rotating magnetized
stars.
That is,
(1) a spherical grid (R,$\theta$,$\phi$) was
used which assures the high
spatial resolution near the star; the
grid $N_R\times N_\theta = 131\times51$ was used in most of
simulations;
(2) a Godunov-type numerical method was used; the
code was improved compared to RUKL so that now we were able to
deal with extremely fast
rotating magnetized stars (see details in
Ustyugova et al. 2004);
(3) ``quiescent" initial conditions
were used so that we were able to investigate the slow viscous
accretion of disk matter to the star.

We use the same scaling parameters as in RUKL.
The length-scale
$R_0$ is $\approx 2.86
~R_*$, where $R_*$ is the
radius of the star.
The velocity scale is $v_0=(GM/R_0)^{1/2}$,
and the time-scale is $t_0=R_0/v_0$.
The unit of time in our plots is
the rotational period at $r=R_0$: $P_0=2\pi t_0$.
The other
scaling parameters are those for the magnetic field, $B_0$,
the density, $\rho_0$, and the pressure, $p_0$.

The corotation radius,
$r_{cr}$, is the distance at which
the angular velocity of the star $\Omega_*$ equals
the Keplerian angular velocity of the disk,
$r_{cr}=(GM/{\Omega_*}^2)^{1/3}$.
In dimensionless form, $\tilde
r_{cr}=1/{\tilde\Omega_*}^{2/3}$,
where $\tilde\Omega_*=\Omega_*/\Omega_0$, and
$\Omega_0=(GM/R_0^3)^{1/2}$.
We use the dimensionless
angular velocity normalized to the Keplerian velocity at the surface
of the star: $\omega_*=\Omega_*/\Omega_{*K}=(R_*/r_{cr})^{3/2}$.
Subsequently, we use only dimensionless variables $\tilde{R}
=R/R_0$, etc. but drop tildes.
Results obtained in dimensionless form are general
and can be applied to
objects with widely different scales.

The initial conditions are similar to RUKL. Namely, there is a
cold dense disk with inner radius $R_d$ and a hot rarefied corona
elsewhere.
The region $r < R_d$ rotates with the
angular velocity of the star, while in the region $r \ge R_d$
both the disk and the corona rotate with the angular velocity
$\omega=\omega(r)=\omega_K(r)$ so as to avoid an initial
discontinuity of the magnetic field at the boundary between the
disk and corona.
The density and temperature distribution in the
disk and corona are derived from the balance of the gravitational,
centrifugal, and pressure gradient forces. The disk and corona are in
pressure balance.

The initial density distribution
of the disk is different from a stationary,
non-magnetized Shakura-Sunyaev (1973) disk
in the respect that the initial surface density
increases gradually with distance in
the outer part of the disk well outside
of the Alfv\'en radius.
An $\alpha-$type viscosity
was incorporated in our code so
as to give a controllable inflow of matter
from the outer disk.
We tried different $\alpha$ values
in the range $0.01 - 0.05$ and used
$\alpha=0.02$ as the base
one for most of simulations.
As a result of the viscosity
the outer disk evolves slowly towards
a stationary Shakura-Sunyaev disk,
but there is insufficient time
for the equilibrium to be reached.
The conditions at the
outer boundary (located at $R_{max}\approx 27$)
allow the inflow of additional
matter to the computational region.
However, matter inflow across $R_{max}$
was not significant for the simulation
times discussed here.

Compared with RUKL, we gradually
increased the rotation rate of the star from a
small value $\omega_*=0.014$ up
to the final value $\omega_*=0.58$
during several rotations of the star.
The inner radius of the disk was
placed at the corotation
radius corresponding to the initial
spin, $R_d=r_{cr}=6$, that is, far away from the star.
These modifications of initial
conditions helped to overcome numerical
difficulties in this
extreme case of a rapidly rotating star.
These initial conditions were
tested and checked in multiple runs with different
initial $R_d$, different times
of rotational acceleration, variations of the
matter distribution in the disk and corona, and sizes of
the simulation region.

\begin{figure*}[b]
\epsscale{1.} \plotone{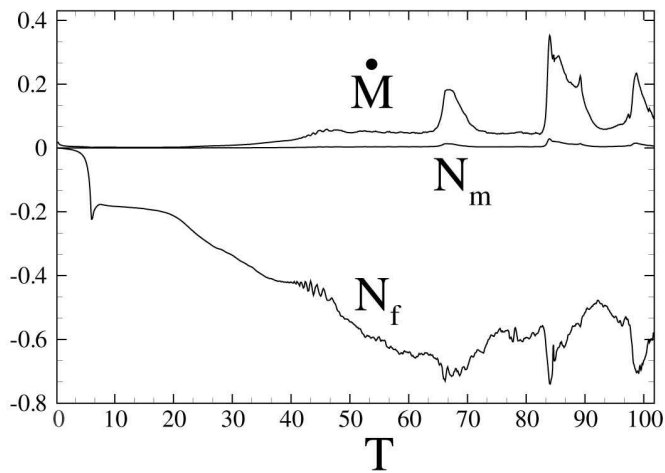} \caption{Temporal evolution of
matter flux to the star $\dot M$, and
torques associated with matter $N_m$ and with magnetic field $N_f$
} \label{Figure 4}
\end{figure*}

\section{Results}

We performed multiple simulation runs for different angular
velocities of the star $\omega_*$ with the other parameters fixed.
The investigated range was from $\omega_*\approx 0.086$
($r_{cr}=1.8$) (weak propeller) to $\omega_*=0.82$ ($r_{cr}=0.4$).
Below we discuss a case where $\omega_*=0.58$ ($r_{cr}=0.5$).

We observed, that initially the disk moved outward so that the
inner radius of the disk increased from $r=R_d=6$ to $r\approx 8
-9$. Then it stabilized around this value. We observed that the
structure of magnetosphere around the star was different from the
magnetospheres of slowly rotating stars. Namely, the magnetosphere
expanded in both radial and vertical directions. Magnetic field
lines from the poles of the star expanded vertically and open,
forming ``magnetic towers" (Figure 1, top). Only a fraction of the
total magnetic flux ($\sim 1/3$) goes into the magnetic tower. The
remainder of the flux forms a {\it closed} magnetosphere, which
does not expand vertically, but instead expands in radial
direction. This is shown in the lower panel of Figure 1. The
magnetic field is strongly twisted in this magnetosphere so that
the toroidal component is much larger than the other components.
This is shown in the left-hand panel of Figure 2. The field varies
as $B_\phi\sim r^{-0.85}$ in the equatorial region. Note that
similar gradual distribution of the field was found in case of
spherical accretion to the propelling star (Romanova et al.
2003a).

The magnetosphere expands due to the rapid rotation of the star.
The accreting matter stops at the Alfv\'en radius $r_A$, where the
matter pressure equals the magnetic pressure, $\rho v_\phi^2 + p
\approx B^2/8\pi$ . Because the $B-$field becomes non-dipolar, we
can not use the standard formulae for $r_A$, but instead must to
find $r_A$ numerically. We found that the Alfv\'en radius
$r_A\approx 8-9$ which is much larger than that for a non-rotating
star ($r_A\approx 1$). We performed simulations at a wide variety
of $\omega_*$ and found that the Alfv\'en radius increases with
$\omega_*$ as $r_A\sim \omega_*$.

The magnetosphere rotates with a super-Keplerian velocity, but not
rigidly as shown in the right-hand panel of Figure 2. The angular
velocity gradually decreases from a maximum value near the star to
a Keplerian velocity of the disk at the magnetospheric radius
$r_A$. Thus, there is no sharp transition in angular velocity
between the magnetosphere and the disk and ``classical" propeller
effect of pushing away incoming matter by the rigidly rotating
magnetosphere is not observed. Instead, the disk ``stands off"
near $r_A$ and tends to be ``lifted'' out of the equatorial plane.
This leads to the formation of elongated funnel stream (see Figure
3, left panel). This stream first encounters the barrier formed by
the rotating magnetic tower. However, the stream gradually forces
the magnetic field lines of the tower to reconnect and thus
creates the path for accretion to the surface of the star (middle
panel). After the accretion event, the magnetic field of the tower
expands, forming an obstacle to further accretion (right panel).
Then matter accumulates again and a new cycle of the accretion
begins (see animation in:
$www.astro.cornell.edu/us-rus/disk\_prop$). This phenomenon of
quasi-periodic outbursts associated with reconnection of the
magnetic field lines was predicted by Aly \& Kuijpers (1990) and
was observed in numerical simulations of accretion to slowly
rotating stars (see Goodson et al. 1997, 1999; RUKL; van Rekowsky
\& Brandenburg 2004; Kato, Hayashi \& Matsumoto 2004).

Accretion is possible because in the magnetosphere of the
propelling star the angular velocity decreases with the distance.
At the distance $r\approx r_A$, the centrifugal force is about
equal to the gravitational force. This is why matter accretes
through the funnel streams as in cases of slowly rotating stars.
In case of propeller the variability is much stronger because the
magnetic tower temporarily blocks the accretion. The quasi-period
observed in the case of the propeller with $\omega_*=0.58$ is
about $\Delta T\approx 15-17$ is large compared to cases of
propellers with smaller angular velocity, $\Delta T\approx 3-5$
(RUKL).

We estimated the accretion rate through the disk at the distance
$r=10$ which is beyond the Alfv\'en radius. We observed that the
inner regions of the disk oscillate in the radial direction with a
quasi-period $\Delta T \approx 5-6$ which is shorter than the main
quasi-period of accretion to the star. The average accretion rate
is $\dot M \approx 0.1-0.2$.

Angular momentum is transported between the star and the disk by
the magnetic field of the closed magnetosphere. The magnetic tower
influences to the dynamics of gas flow around the propelling star,
but it is not important for the angular momentum transport.

Figure 4 shows the temporal evolution of the matter flux to the star
$\dot M$, and the torques due to the
matter flow ${\bf N}_m$, and due
to the magnetic field ${\bf N}_f$.
The total torque is mainly due to the magnetic field. The torque
due to the accreting matter is negligibly small even in periods of
enhanced accretion, like in cases of slowly rotating stars (RUKL).
We note that the star spins down {\it continuously}, even during
periods when the accretion is blocked by the magnetic tower. When
the accretion rate increases, the {\it spin-down torque also
increases}. This is in opposite to the standard spin-down theory,
where the spin-down torque decreases with accretion rate. The
difference is due to the fact that during periods of enhanced
accretion, a larger fraction of the star's flux is in the closed
magnetosphere. This leads to a larger spin-down torque. Running a
number of cases with different $\omega_*$, we obtained the
dependence for the average torque in the form $\dot N_f \sim -
\omega_*^{1.3}$. Note, that the same dependence was found in
simulations of quasi-spherical accretion to the propelling star,
which were done with  different numerical code (Romanova et al.
2003a).

The early ideas that accreting stars may spin-down due to the
magnetic disk-star interaction were developed by Ghosh \& Lamb
(1979) and included the prediction that a star may accrete and
spin-down at the same time. This last conclusion was observed in
RUKL for the case of a weak propeller. The effect is observed in
the present simulations of much more rapidly rotating stars. The
spin-down rate is observed to {\it increase} with accretion rate.
This dependence is new and different from that for slowly rotating
stars.

\section{Discussion}

We observed that disk accretion to a star in the propeller regime
exhibits a number of interesting new features: (1) The most
remarkable one is the fact that the {\it rate of spin-down
increases as the accretion rate increases.} This is opposite the
dependence in the non-propeller regime (e.g., Ghosh \& Lamb 1979).
(2) The fact that accretion {\it occurs} in the strong propeller
regime is also important. The accretion is {\it quasi-periodic}
and occurs through elongated funnel streams which go around the
region of centrifugally-dominated magnetosphere. (3) About 1/3 of
the star's magnetic flux goes to the magnetic tower which
represents a region of opened field lines. However, the remainder
of the flux goes into a radially expanded closed magnetosphere,
which connects the star and the disk. This magnetic field is
responsible for the strong angular momentum transport between the
star and the disk. (4) The propeller stage may be very efficient
in spinning-down rapidly rotating accreting stars.

It was recently suggested that the opening of the magnetic field
lines disconnects the star and the disk (e.g., Lovelace et al.
1995) so that the spin-down of the star should be significantly
reduced (Agapitou \& Papaloizou 2000; Matt \& Pudritz 2004). Our
simulations show that most of the magnetic flux of the star is in
a closed magnetosphere and is coupled to the accretion disk. The
angular momentum of the star is magnetically transported to the
disk. Thus, the propeller mechanism may explain the spinning-down
of the Classical T Tauri stars.

For much stronger magnetic fields, propeller-driven outflows were
observed, which will be discussed in a subsequent paper (Ustyugova
et al. 2004).  Recently, we were able to perform full 3D
simulations of the disk accretion to an inclined dipole (Koldoba
et al. 2002, Romanova et al. 2003b, 2004). Next, we plan to
investigate the propeller regime using fully three-dimensional MHD
simulations.

\acknowledgments This work was supported in part by NASA grants
NAG5-13220, NAG5-13060, and by NSF grant AST-0307817. AVK and GVU
were partially supported by grants INTAS CALL2000-491, RFBR
03-02-16548, by contract MEST \# 40.022.1.1.1106 and by Russian
program ``Astronomy''. Authors are grateful to anonymous referee
for valuable comments and suggestions.

\end{document}